\documentclass{emulateapj}
\usepackage{apjfonts}

\input psfig.sty

\slugcomment{ApJ, in press}

\shorttitle{Eccentricity of supermassive black hole binaries}
\shortauthors{Armitage \& Natarajan}

\begin{document}

\title{Eccentricity of supermassive black hole binaries \\ coalescing from gas rich mergers}

\author{Philip J. Armitage\altaffilmark{1,2} and Priyamvada Natarajan\altaffilmark{3,4}}

\altaffiltext{1}{JILA, Campus Box 440, University of Colorado, Boulder CO 80309;  
pja@jilau1.colorado.edu}
\altaffiltext{2}{Department of Astrophysical and Planetary Sciences, University
of Colorado, Boulder CO 80309}
\altaffiltext{3}{Department of Astronomy, Yale University, P.O. Box 208101, 
       New Haven, CT 06520-8101; priya@astro.yale.edu}
\altaffiltext{4}{Department of Physics, Yale University, P.O. Box 208120, 
New Haven, CT 06520-8120}

\begin{abstract}
Angular momentum loss to circumbinary gas provides a possible
mechanism for overcoming the `last parsec' problem and allowing
the most massive black hole binaries formed from galactic mergers to
coalesce.  Here, we show that if gas disks also catalyze the merger of
the somewhat lower mass binaries detectable with the {\it Laser
Interferometer Space Antenna} (LISA), then there may be a purely
gravitational wave signature of the role of gas in the form of a small
but finite eccentricity just prior to merger. Numerical 
simulations suggest that eccentricity, excited by the interaction between 
the binary and surrounding gas disk, is only partially damped during the 
final phase of gravitational radiation-driven inspiral. We estimate 
a typical eccentricity at one week prior to coalescence of 
$e \approx 0.01$. Higher terminal eccentricities, which can 
approach $e = 0.1$, are possible if the binary has an extreme 
mass ratio. The detection of even a small
eccentricity prior to merger by LISA provides a possible discriminant 
between gas-driven inspirals and those effected by stellar
processes.
\end{abstract}


\section{Introduction}
Supermassive black hole binaries formed from galactic mergers can coalesce 
if angular momentum loss -- initially to stars and gas and subsequently 
via gravitational radiation \citep{begelman80} -- is fast enough to 
conclude the process within at most a Hubble time, or more stringently 
before a subsequent merger creates a 3-body system that is  
unstable to a black hole ejection \citep{valtonen94}. Due to the formation of a 
loss cone, stars on their own are likely to fail to meet this requirement 
for binary separations of the order of $r \sim 1$~pc and binary masses
$M \gtrsim 10^8 \ M_\odot$ \citep{milos01,yu2002,berczik05}, parameters which today are 
characteristic of massive ($\sigma \sim 200 \ {\rm km \ s^{-1}}$) galaxies  
\citep{gebhardt2000,ferrarese2000}. For this subset of galaxies, angular 
momentum loss to gas provides a plausible mechanism that can yield a 
prompt merger \citep{begelman80,gould2000,phil2002,escala04,escala05}.  
At lower masses, stellar dynamic processes are in principle 
sufficient to effect mergers, but gas may still be involved if disks 
are present and remove angular momentum efficiently on a time scale 
significantly shorter than stellar interactions.

Gas disks that catalyze binary mergers may generate observable 
electromagnetic counterparts to the primary gravitational wave signature 
detectable with the {\em Laser Interferometer Space Antenna} (LISA). 
Induced accretion or outflows from small accretion disks surrounding 
the individual holes might yield precursors to the gravitational wave 
event \citep{phil2002}, while the inflow of circumbinary gas left 
stranded as the binary contracts under gravitational radiation losses 
provides a robust prediction of an 
afterglow \citep{milos05}. On a longer time scale, impulsive changes 
to the spin of the hole following merger \citep{hughes03} are likely 
to change the direction of any jets launched from the immediate 
vicinity of the black hole \citep{merritt02}. Such counterparts are 
of interest, in part, because their detection in coincidence with LISA 
events would overcome LISA's limited capability to spatially localize 
signals, and pin down the galaxy within which a merger had occurred.
Unambiguous identification of merger sites significantly enhances the 
potential cosmological utility of merger events \citep{holz05,kocsis05}, 
and substantially improves the ability of LISA to trace the growth of 
black holes as a function of galactic environment and morphology.

In this paper, we study the evolution of binary eccentricity in the 
event that angular momentum loss to gas dominates the inspiral 
immediately prior to the onset of rapid shrinkage via gravitational 
radiation. In \S2 we show that the gravitational interaction between 
the binary and a surrounding gas disk is likely to excite black 
hole binaries to eccentricities $e > 0.1$, a result anticipated from 
both analytic studies \citep{artymowicz92,goldreich03}, simulations  
of pre-main sequence binary stars \citep{artymowicz91}, and simulations of 
massive disk-embedded 
planets \citep{papaloizou01}. In \S3 we argue that although gravitational 
radiation will damp these eccentricities prior to coalescence, the 
transition between disk-driven and gravitational wave-driven inspiral 
can occur at small enough radii that a small but significant 
eccentricity survives when the separation is only tens of 
gravitational radii. Detection of this eccentricity in the gravitational 
wave signal would provide strong circumstantial evidence for the role 
of gas in driving mergers, even in the absence of other
electromagnetic counterparts.

\section{Eccentricity during gas-driven inspirals}
If gas is the causative angular momentum loss process of supermassive 
black hole binary coalescences, the merger is expected to proceed 
through two phases. In the initial phase, the binary is either 
surrounded by \citep{gould2000,phil2002} or completely embedded 
within \citep{escala04} circumbinary gas. If the gas mass is not 
too large, and the gas temperature (relative to the virial temperature 
at the radius of the binary) not too high, gravitational torques \citep{goldreich80} 
from the binary act to clear a gap in the circumbinary disk \citep{lin80}. 
The inner truncation radius of this gap depends upon the binary 
mass ratio, disk properties, and eccentricity. For small eccentricity 
truncation occurs at around $2a$, where $a$ is the binary semi-major axis 
\citep{artymowicz94}. The 
binary, and the inner edge of the circumbinary disk, shrink as 
angular momentum is transfered from the orbital angular momentum 
of the black holes into the gas. This phase is qualitatively 
similar to the Type~II migration (i.e. involving a gap) of 
massive planets, with the major difference being that a 
close black hole binary interacts with a relatively smaller mass 
of disk gas than a typical massive planet. The mismatch between 
the disk mass and the binary mass slows the rate of both gas 
inflow and orbital decay.

The first phase ends when angular momentum loss to gravitational 
radiation -- which on its own drives decay of the semi-major 
axis $a$ at a rate $\dot{a} \propto a^{-3}$ -- first exceeds 
that due to circumbinary disk torques. The binary then enters 
a second phase in which gravitational wave losses control the 
rate of inspiral. Initially, the circumbinary gas, although 
now dynamically unimportant, flows inward  
fast enough to remain in contact with the tidal barrier 
created by the binary. Any initial eccentricity excited 
during the first phase will start to damp. Eventually, however, the decay 
of the binary separation becomes too fast, and the binary 
detaches from the circumbinary gas, which through coalescence remains 
approximately frozen in the configuration it had at the moment of detachment.  
\citep{phil2002,milos05}. Any low density gas remaining within the 
central cavity during this 
final stage is expected to have little influence on the binary orbital 
elements \citep{narayan00}. 

\subsection{Numerical simulations}
We have simulated the interaction between a binary and surrounding 
circumbinary gas disk using the Zeus code \citep{stone92}. We work 
in 2-dimensional cylindrical co-ordinates $(r,\phi)$, with `scaled 
gridding' in the radial direction (i.e. $r_{i+1} = (1+\epsilon) r_i$, 
with $\epsilon$ a constant, so that computational 
zones have the same shape at all radii) and a 
standard resolution of $n_r = 400$, $n_\phi = 400$. The computational 
domain extends from $r_{\rm in}$ to $r_{\rm out}$, with $r_{\rm in}$ 
chosen to lie interior to the expected inner truncation radius 
of the circumbinary disk \citep{artymowicz94}. We exclude the 
region close to the black holes, and do not attempt to model 
the gas flow or torques arising from the smaller disks around 
each black hole. Outflow boundary conditions are imposed at both 
$r_{\rm in}$ and $r_{\rm out}$.

For a thin disk, the geometric thickness $h/r$ is related to the 
local sound speed $c_s$ via $h/r \simeq c_s / v_K$, where $v_K$ 
is the Keplerian orbital velocity about the binary's center of 
mass. We consider self-similar disk models in which the equation 
of state is isothermal at any radius ($P=\rho c_s^2$), with an 
imposed radial scaling,
\begin{equation}
 c_s(r) \propto r^{-1/2}.
\end{equation}
We adopt units in which $G=1$, the sum of the masses of the 
black holes $M_1 + M_2 = 1$, and consider models in which 
$h/r = 0.2$ (corresponding to $c_s = 0.2$ at $r=1$), 0.1 and 0.05.  
We refer to these as the `hot', `warm' and `cold' disk 
models respectively.

At the relatively small radii we are most interested in ($r \lesssim
10^3 r_s$, where $r_s$ is the Schwarzschild radius), angular 
momentum transport within the circumbinary disk
probably arises from turbulence driven by the Magnetorotational
Instability (MRI) \citep{balbus91}, which cannot be simulated in two
dimensions.  Instead we model angular momentum transport within the
disk using a constant kinematic viscosity $\nu$, which operates on the
azimuthal component of the momentum equation only
\citep{papaloizou86}. There is no reason to think that this, or any
other simple viscosity prescription, reproduces accurately the
multi-dimensional effects of angular momentum transport from MHD
turbulence \citep{ogilvie03b}. We adopt it purely for numerical
simplicity. The magnitude of $\nu$ is set by reference to the
\citet{shakura73} viscosity prescription,
\begin{equation}
 \nu = { {\alpha c_s^2} \over \Omega}.
\end{equation}
We choose $\nu$ such that the equivalent $\alpha=0.1$ at 
$r = 2a$, where $a$ is the initial binary separation.
Consistent with the choice of a constant viscosity, we set the 
disk surface density $\Sigma(r)$ to be uniform in the initial 
state, so that the initial accretion rate $\dot{M}(r) \propto \nu \Sigma$ 
is constant with radius. The disk models with varying $h/r$ are 
set up with the same initial surface density.

The dynamics of the binary, including forces from the circumbinary 
gas, are integrated using a simple second-order scheme that is 
adequate for the relatively short duration ($t \sim 10^2$ orbits) 
of our simulations. Over these time scales, realistic surface 
densities of the circumbinary disk are small enough that they 
would lead to negligible 
changes in the binary semi-major axis $a$ and eccentricity $e$. 
For example, a steady-state Shakura-Sunyaev disk \citep{shakura73} 
around a black hole of mass $10^6 \ M_\odot$, accreting at the 
Eddington limit with $\alpha = 0.1$, has a surface density 
at $10^3 r_s$ of the order of $10^5 \ {\rm g cm^{-2}}$. This 
means that the mass in the region of the disk that interacts 
with the binary is only a very small fraction of 
the binary mass, and evolution is slow.  
In order to see significant evolution within of the order of 
$10^2$ orbits, we therefore need to adopt 
a higher scaling for the surface density. 
However, this {\em does not} affect the rate of eccentricity growth 
(or damping), measured as $a {\rm d}e / {\rm d}a$, because the 
perturbing accelerations that alter $a$ and $e$ are linear in the 
disk mass (this is evident explicitly in Gauss' equations, given 
in simplified form as equations (1)-(3) of Artymowicz et al. 1991).
Choosing a higher surface density scaling therefore accelerates 
the time scale for changes in both $a$ and $e$ by the same factor. 
Large disk masses do lead to some wandering of the binary as 
the eccentricity grows -- we suppress this and maintain the 
binary center of mass at $r=0$ throughout the runs.

The initial conditions are set up as a uniform disk in 
Keplerian rotation (including the correction due to the 
pressure gradient) outside $r = 2a$, where $a$ is the initial 
binary separation. Inside $2a$, we exponentially truncate the 
surface density. For the hot disk run, we evolve the disk 
for 40 binary orbits before turning on (smoothly over an 
additional 5 orbits) the backreaction from the disk in the 
binary integration. For the cooler disks, we start with the 
initial conditions for the hot disk and linearly reduce 
the sound speed over the first 25 orbits of the calculation 
until it reaches its final, fixed value. These procedures 
are intended to minimize {\em dynamical time scale} transients 
created as the initially uniform disk adjusts to the non-axisymmetric 
forcing from the binary. The inner edge of the circumbinary 
disk, for example, is able to adjust its position in 
response to the forcing on such a short time scale. Considering 
longer time scales, however, the initial disk surface density 
profile -- which corresponds to a steady-state solution around 
a single central mass -- is not in equilibrium. Over a viscous 
time scale, gas whose inflow is impeded by tidal forces from the 
binary would accumulate near the inner edge of the circumbinary disk, 
increasing the surface density. Over the duration of our 
simulations, this effect is small but noticeable. It is 
not present for the hottest disk, which is able to overflow 
the gap toward the individual binary components most easily.

\subsection{Results}

\begin{figure}
\psfig{figure=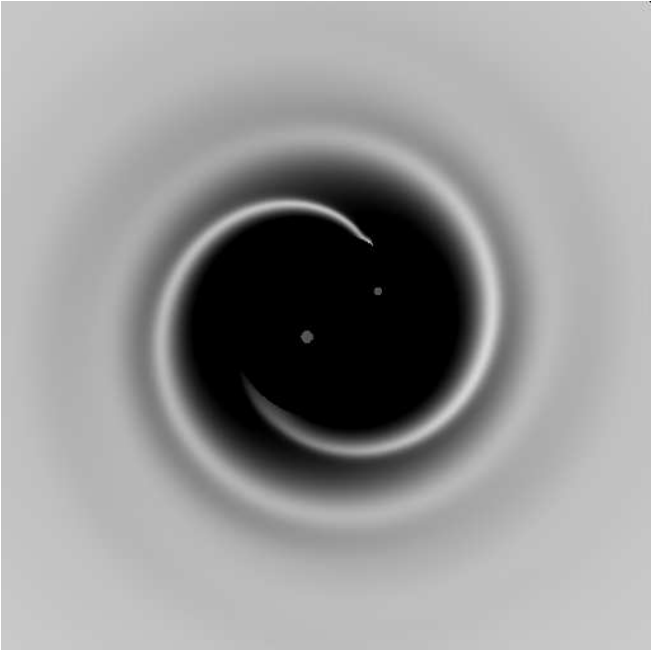,width=\columnwidth}
\figcaption{Surface density at $t=40$~orbits, just prior to starting active integration 
of the binary, from the hottest disk with $h/r = 0.2$. The image is $8a$ on 
a side, where $a$ is the initial binary separation. For this run the mass ratio 
$q \equiv M_2 / M_1 = 1/3$, and the initial eccentricity was $e=0.02$.}
\label{fig_image}
\end{figure}

Figure~1 shows a map of the surface density at $t = 40$~orbits 
for the hot disk, which surrounds a binary with mass ratio 
$q = M_2 / M_1 = 1/3$, initial semi-major axis $a=1.02$, and 
initial eccentricity $e=0.02$. The 
computational domain extended from $r_{\rm in} = 1.15$ to 
$r_{\rm out} = 20$. Torques from the binary truncate the 
inner edge of the circumbinary disk at around $2a$, although 
there is significant overflow across the gap \citep{artymowicz96} 
and through the inner boundary of the simulated disk. The 
cooler disk runs are qualitatively similar, although the 
thickness of the spiral arms and the extent of mass overflow 
both decrease as the disk becomes thinner. 

\begin{figure}
\psfig{figure=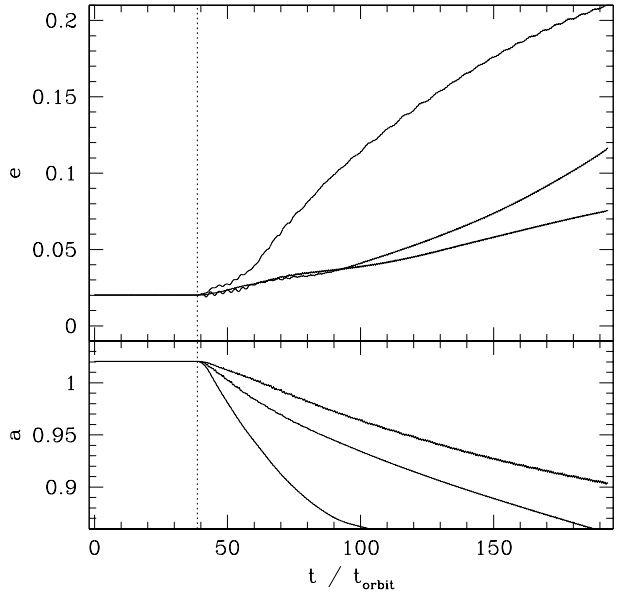,width=\columnwidth}
\figcaption{Evolution of the binary eccentricity (upper panel) and semi-major 
axis (lower panel) for simulations of binaries embedded within disks of 
varying geometric thickness. The eccentricity has been averaged over the 
orbital period to filter out small short time scale variations.
The rate of eccentricity growth increases with 
decreasing disk thickness, while the rate of decay of the separation is reduced.
Note that the normalization of the disk surface density 
has been chosen (arbitrarily) to yield measurable evolution over the 
$\sim 100$ orbit time span of the simulations.}
\label{fig_growth}
\end{figure}

Figure~2 shows the evolution of $a$ and $e$ for binaries 
embedded within the cold, warm, and hot disks. Once the 
backreaction forces from the disk are turned on after the 
initial equilibration period, the eccentricity grows 
and the orbit decays in all three cases. As expected, 
the rate of orbital decay is fastest for the hottest, 
most viscous disk model, while eccentricity grows most 
rapidly for the coldest disk. This is consistent with 
the analytic expectation that eccentricity growth is 
driven by resonant interaction with gas at relatively 
large distance from the binary. For all three of our 
disk models, we find that the inner edge of the disk 
at the end of the equilibration period remains close to 
$r \simeq 2a$. The dominant active resonance is then 
the 1:3 outer eccentric Lindblad resonance at $r = 2.08a$ 
\citep{artymowicz94}. The relative importance 
of this resonance is increased for cold disks, in 
which the inner evacuated cavity is better 
defined. For the cold disk model, although we are unable 
to determine the saturation value of the eccentricity, 
we find that $e$ increases to more than  $e=0.1$ before 
the binary has shrunk by more than a few percent in 
semi-major axis. 

\begin{figure}
\psfig{figure=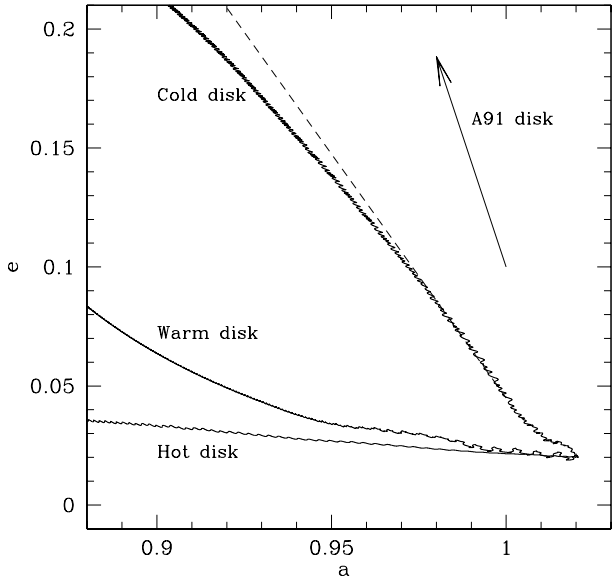,width=\columnwidth}
\figcaption{Binary evolution in the $a$-$e$ plane for disks of varying geometric 
thickness. The highest initial rates of eccentricity growth 
($a {\rm d}e / {\rm d}a \simeq -2$), shown as the dashed line) 
are obtained for the coolest, thinnest disk simulated. The arrow shows the estimated 
growth rate for the even colder disk model simulated by Artymowicz et al. (1991).}
\label{fig_deda}
\end{figure}

Figure~3 shows the evolution of binary orbital elements in 
the $a$-$e$ plane, which for the reasons noted above is 
independent of the arbitrary choice of disk surface 
density. For the cold disk we find that the eccentricity 
grows at a relatively constant rate,
\begin{equation} 
 a { {{\rm d}e} \over {{\rm d}a} } \simeq -2.05,
\end{equation}
up to $e \approx 0.1$, with the growth rate modestly 
decreasing as the eccentricity increases further. Any 
significant orbital migration within such a disk would 
therefore be expected to yield significant eccentricities. 
Growth rates are substantially lower in the warm 
($a {\rm d}e / {\rm d}a \sim -0.25$) and cold 
($a {\rm d}e / {\rm d}a \approx -0.1$) disks. This trend 
is qualitatively consistent with how one might expect  
${\rm d}a / {\rm d}t$ and ${\rm d}e / {\rm d}t$ to scale with binary 
parameters. Defining the ratio of the local disk mass to the 
secondary mass via $q_d = \pi a^2 \Sigma / M_2$, we expect 
the disk interaction to shrink the binary orbit at a rate,
\begin{equation}
 { {{\rm d}a} \over {{\rm d}t} } \propto - \alpha \left( {h \over r} \right)^2 
 q_d \Omega a. 
\end{equation} 
This is just the viscous inflow velocity of gas in the disk, reduced 
by the ratio of the local disk mass to the secondary mass. We note 
that this expression does not include changes to the equilibrium 
disk structure that occur over a viscous time scale \citep{syer95}. 
Viscous time scale equilibrium, which is not achieved in our simulations, 
is discussed more thoroughly in \S3. The rate of 
change of eccentricity is expected to scale as \citep{goldreich80},
\begin{equation}
 { {{\rm d}e} \over {{\rm d}t} } \propto q^2 q_d \Omega e.
\end{equation}
Comparing these expressions we expect eccentricity growth to be most 
rapid for nearly equal mass ratio binaries in geometrically thin and / or 
relatively inviscid disks.

\subsection{Comparison with previous work}

Our simulations are similar in spirit to those of \citet{artymowicz91} and 
\citet{gunther2004}. Using smooth particle hydrodynamics simulations, \citet{artymowicz91} 
derived an eccentricity growth rate $a {\rm d}e / {\rm d}a \approx -4.4$ 
for a binary with a mass ratio of $q=3/7$ and $e=0.1$, embedded within a disk 
in which $h/r \approx 0.03$ and $\alpha \sim 0.1$. This disk 
is modestly colder than our `cold' case. Their growth 
rate, shown as an arrow in Figure~3, is approximately 
a factor of two greater than obtained in our `cold' simulation. Our 
smaller growth rate is consistent with a trend toward faster growth in 
geometrically thinner disks. 

Motivated by the surprisingly large eccentricities of many massive 
extrasolar planets, more recent numerical work on binary-disk 
interactions has focused on extreme mass ratio ($q \ll 1$) systems. 
In a disk model intended to represent a typical protoplanetary 
disk, \citet{papaloizou01} obtained eccentricity growth for 
mass ratios $q \gtrsim 0.02$, while damping occurred for more 
extreme mass ratios. Eccentricities as large as $e \approx 0.25$ were 
obtained in these simulations. Subsequent analytic work \citep{ogilvie03,goldreich03} 
has identified circumstances in which the critical mass ratio for 
eccentricity growth might be reduced still further -- perhaps 
to values of $q \sim 10^{-3}$  -- but this has yet to be confirmed 
numerically. In the black hole binary problem, these subtleties 
are probably moot. Extreme mass 
ratios are unlikely to occur as a consequence of galactic mergers 
\citep{yu2002}, because the dynamical friction time scale for the 
smaller black hole to reach the galactic center would exceed the 
Hubble time. Extreme mass ratio binaries are however conceivable if 
intermediate mass black holes \citep{miller04} form within the 
nucleus and are subsequently captured by the gas disk. 

\subsection{Initial eccentricity}

Our simulations commence with finite eccentricity, which is 
then unstable to subsequent growth. The physical origin 
of such a seed eccentricity is speculative, although there are 
enough possibilities that we think it unlikely that a 
binary could evade the instability by remaining perfectly 
circular. First, the endpoint of the prior stage of 
stellar dynamical hardening might leave the binary with 
non-zero eccentricity, although the results of 
\citet{berczik05} suggest that any such eccentricity would 
be small. Second, the binary could acquire eccentricity 
while embedded within the disk, for example via a Kozai 
resonance with a misaligned gas or stellar disk at 
larger radius in the nucleus \citep{blaes02}. Finally, 
non-axisymmetry in the disk itself -- due to self-gravity 
at large radii \citep{goodman03}, eccentric instabilities 
in individual circumstellar disks \citep{lubow91}, or 
instabilities at the edge of the circumbinary disk 
\citep{goldreich03} -- appears to be almost unavoidable. 
Such non-axisymmetry in the disk is readily exchanged 
into the binary \citep{papaloizou01,papaloizou02}, and, 
if it is large enough, could serve as the initial seed 
for subsequent growth.

\section{Eccentricity of binaries close to coalescence}

Any eccentricity attained by the binary during the phase of 
gas-driven orbital decay will be damped during the final 
stage of gravitational wave-driven inspiral. The extent of 
the damping depends upon the separation of the binary (in 
gravitational radii) at the time when decoupling from the 
gas occurs. To gain a rough idea of the magnitude of the 
damping, we initially consider a toy model in which the 
binary decouples instantaneously from the disk at a radius 
$a_{\rm crit}$, which for 
now we regard as a free parameter. For $a < a_{\rm crit}$, 
the binary orbit evolves solely 
under the influence of gravitational radiation losses. For a 
binary with energy $E$ and angular momentum $J$, at such a 
stage \citep{padmanabhan1},
\begin{eqnarray}
\frac{dE}{dt} &=& -\frac{32G^4\,M_1^{2}\,M_2^{2}(M_1+M_2)}{5\,c^5\,a^5}\,\times\,\frac{1}{(1
  - e^2)^{\frac{7}{2}}} \nonumber \\ 
 && \times \, [1 + \frac{73}{24}\,e^2 + \frac{37}{96}\,e^4], \\
\frac{dJ}{dt} &=& -{\frac{32\,G^{\frac{7}{2}}\,M_1^2\,M_2^2\,{\sqrt{M_1
    + M_2}}}{5\,c^5\,a^{\frac{7}{2}}}}\,\times\,{\frac{1}{(1 -
    e^2)^2}} \nonumber \\
    && \times \, [1 + \frac{7}{4}\,e^2].
\end{eqnarray}
These equations are readily integrated to yield the residual eccentricity 
at a fiducial time (typically either one week or one year) prior to coalescence, 
as a function of $a_{\rm crit}$, $q$, and the assumed initial eccentricity 
generated during the earlier gas-dominated phase. The formulae 
apply in the weak field limit, which is appropriate 
provided that we do not consider times {\em extremely} close to 
the instant of coalescence (note that at one week before coalescence, 
the separation is still $a \approx 40 GM / c^2$). In the strong 
field regime, $e$ tends to increase in anticipation of the final 
`plunge' \citep{cutler94}.

\begin{figure}
\psfig{figure=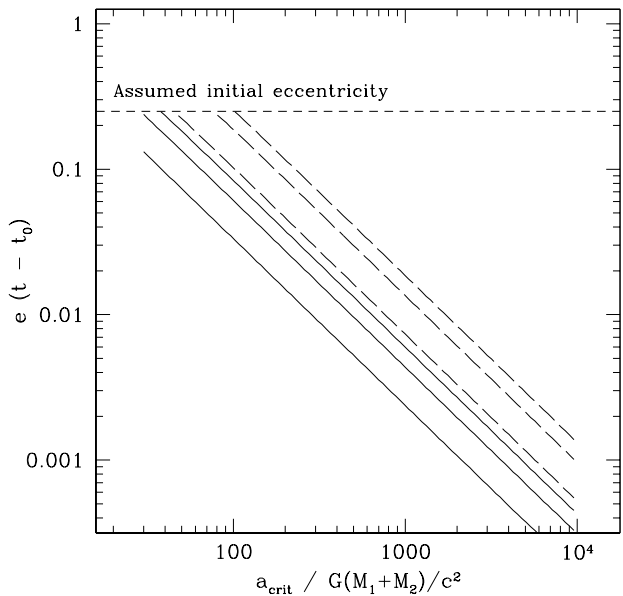,width=\columnwidth}
\figcaption{Binary eccentricity at a time $t_0$ prior to coalescence in a toy model 
in which the binary is assumed to attain an eccentricity of $e=0.25$ during the 
initial phase of gas-driven inspiral at $a > a_{\rm crit}$. At $a_{\rm crit}$ we 
assume that gravitational radiation starts to dominate over angular momentum loss 
to the gas, so that for $a < a_{\rm crit}$ inspiral and attendant damping of $e$ 
is governed solely by gravitational radiation losses. The solid lines show 
results for binary mass ratios $q \equiv M_2 / M_1$ of 1, 0.1, 0.01 
(the upper solid curve is for $q=1$), and $t_0 = \ {\rm 1 \ week}$. The dashed 
curves are for $t_0 = \ {\rm 1 \ year}$.  
The total system mass is $10^6 \ M_\odot$ in all cases.}
\label{fig_damp}
\end{figure}

Figure~4 shows the predicted binary eccentricity at one week and one year before 
coalescence for different $q$, as a function of the decoupling radius 
$a_{\rm crit}$. An initial eccentricity $e=0.25$ was assumed, which 
seems reasonable given our simulations and previous work 
\citep{artymowicz91,artymowicz92,papaloizou01}. Binaries observed 
through coalescence with LISA are predicted to have a mass ratio 
distribution that peaks around $q \simeq 0.1$. If these binaries 
in fact have a gas-driven inspiral phase, decoupling within $10^3$ 
gravitational radii typically yields a final eccentricity of 
$e \sim 0.01$. At fixed $a_{\rm crit}$, binaries with more 
extreme mass ratios yield smaller final values of the 
eccentricity, reflecting the fact that at a given time prior 
to coalescence the separation of $q \ll 1$ systems is smaller 
than for the $q=1$ case.

To go further, we need to estimate how $a_{\rm crit}$  depends 
upon the binary and disk parameters. We begin by noting that, 
in the absence of a binary, gas in an accretion disk flows 
inward at a rate given by the viscous inflow speed,
\begin{equation} 
 \dot{a}_{\rm visc} = - {3 \over 2} \left( {h \over r} \right)^2 
 \alpha v_K.
\end{equation}
The separation of a circular binary shrinking under the influence 
of gravitational radiation decays at a rate,
\begin{equation}
 \dot{a}_{\rm GW} = - { {64 G^3 M_1 M_2 (M_1 + M_2)} \over 
 {5 c^5 a^3} }.
\end{equation}
Equating these expressions determines a {\em detachment} radius, 
defined via $\dot{a}_{\rm visc} (a_{\rm detach}) = 
\dot{a}_{\rm GW} (a_{\rm detach})$. The detachment radius is 
the smallest radius for which the inflowing gas can `keep up' 
with the shrinkage of the binary. Interior to this radius, the 
binary orbit continues to contract within a growing hole while 
the circumbinary gas remains essentially frozen \citep{phil2002,
milos05}. The detachment radius is typically quite small. For 
example, \cite{milos05} estimate that the shrinking binary 
leaves a hole of size $\approx 120 GM/c^2$, which lies well 
within the innermost, radiation pressure dominated disk. 

For dynamical purposes we are more interested in 
the decoupling radius $a_{\rm crit}$ -- within 
which the rate of angular momentum loss to gravitational 
waves exceeds that due to interaction with the circumbinary 
disk. To calculate this we need to set $\dot{a}_{\rm GW}$ equal to the disk-driven 
inspiral velocity $\dot{a}_{\rm inspiral}$. The inspiral  
velocity equals the viscous inflow speed in the limit where 
the secondary black hole is of low enough mass that it can 
be treated as a test particle within the disk. If the 
unperturbed disk surface density at the orbital radius $r$ 
of the secondary is $\Sigma_0$, test particle behavior will 
occur for $M_2 \lesssim \pi r^2 \Sigma_0$. For more 
massive secondaries, the large reservoir of binary 
angular momentum reduces the inspiral velocity to a 
fraction of the viscous inflow speed. The 
binary then acts as a slowly 
moving dam which impedes the free inflow of gas.  

For black hole binaries the massive secondary limit is 
the relevant one for small enough separations. To estimate 
the inspiral velocity in this regime, we apply the analysis 
of \cite{syer95}. For a disk in which the unperturbed 
surface density at radius $r$ is $\Sigma_0$, \cite{syer95} 
define a measure of disk dominance,
\begin{equation} 
 B \equiv { {4 \pi r^2 \Sigma_0} \over {M_2} }, 
\end{equation} 
where $M_2$ is the secondary mass. Then if the  
unperturbed disk solution can be locally written as,
\begin{equation} 
 \Sigma_0 \propto {\dot{M}}^\beta r^\gamma,
\label{eq_a} 
\end{equation} 
with $\beta > 0$, the inspiral velocity is,
\begin{eqnarray} 
 \dot{a}_{\rm inspiral} & = & \dot{a}_{\rm visc} \,\,\, (B > 1) \nonumber \\
 \dot{a}_{\rm inspiral} & = & \dot{a}_{\rm visc} B^{\beta/(1+\beta)} \,\,\, (B \leq 1).
\label{eq_slowdown} 
\end{eqnarray}  
The correction factor in the last equation accounts for the slowdown 
in orbital decay when the secondary is massive compared to the local 
disk. The slowdown is usually less than the full factor of $B$, because 
gas accumulates just outside the slowly moving tidal barrier created 
by the binary. This pile-up increases  
the torque relative to that obtained from an unperturbed disk solution 
\citep{pringle91}.

The above equations allow us to calculate the decoupling 
radius for a binary given an estimate for $\alpha$ and a model 
for the accretion disk structure. Here, we adopt the simplest 
Shakura-Sunyaev (1973) model. We normalize the accretion rate 
to the Eddington limiting value,
\begin{eqnarray} 
 \dot{m} & \equiv & { \dot{M} \over \dot{M}_{\rm Edd} } \\
 \dot{M}_{\rm Edd} & = & 2.2 \times 10^{-8} 
 \left( {M \over M_\odot} \right) \, M_\odot {\rm yr}^{-1},
\end{eqnarray}
and measure the mass $m \equiv M / M_\odot$ in Solar units. Then 
in the inner, radiation pressure-dominated disk, away from the 
inner boundary \citep{padmanabhan2}, 
\begin{eqnarray}
 {h \over r_s} & = & 7.46 \dot{m}, \\
 \Sigma & = & 0.42 \alpha^{-1} \dot{m}^{-1} (r/r_s)^{3/2} \ {\rm g cm}^{-2} ,
\end{eqnarray}  
whereas at larger radii, where gas pressure dominates and the 
opacity is due to electron scattering,
\begin{eqnarray}
 {h \over r_s} & = & 1.6 \times 10^{-2} \alpha^{-1/10} 
 \dot{m}^{1/5} m^{-1/10} (r/r_s)^{21/20}, \\
 \Sigma & = & 9.7 \times 10^4 \alpha^{-4/5} \dot{m}^{3/5} m^{1/5} (r/r_s)^{-3/5} \ {\rm g cm}^{-2} .
\end{eqnarray}
Here $r_s = 2GM/c^2$ is the Schwarzschild radius, and the 
transition between the two regimes occurs at,
\begin{equation} 
 {r_t \over r_s} = 360 \alpha^{2/21} \dot{m}^{16/21} 
 m^{2/21}.
\label{eq_transition}
\end{equation}
These relations allow us to calculate $\dot{a}_{\rm visc}$ in both 
the radiation pressure and gas pressure dominated regions of the 
disk. In the gas pressure dominated middle disk we can also 
directly estimate $\dot{a}_{\rm inspiral}$. In this region 
$\beta$ in equation~(\ref{eq_a}) 
is 0.6, so the slowdown in the inspiral velocity when the secondary is massive 
scales as $B^{0.375}$. For representative disk and binary parameters 
($M_1 = 9 \times 10^5 \ M_\odot$, $M_2 = 10^5 \ M_\odot$, $\alpha = 10^{-2}$, 
$\dot{m} = 1$) we find that the disk drives inspiral at a rate 
reduced from the viscous rate by about a factor of 50. 
We cannot use the same formula to calculate 
the slowdown in the radiation pressure dominated inner disk, since 
in this region $\beta < 0$. For an estimate, we simply continue to use the 
slowdown factor derived for the middle disk in this region. We note 
that the structure of accretion disks in the innermost radiation-dominated 
regime remains subject to large uncertainties \citep{turner04,turner05}, 
so any statements about this region are at best semi-quantitative.  

\begin{figure}
\psfig{figure=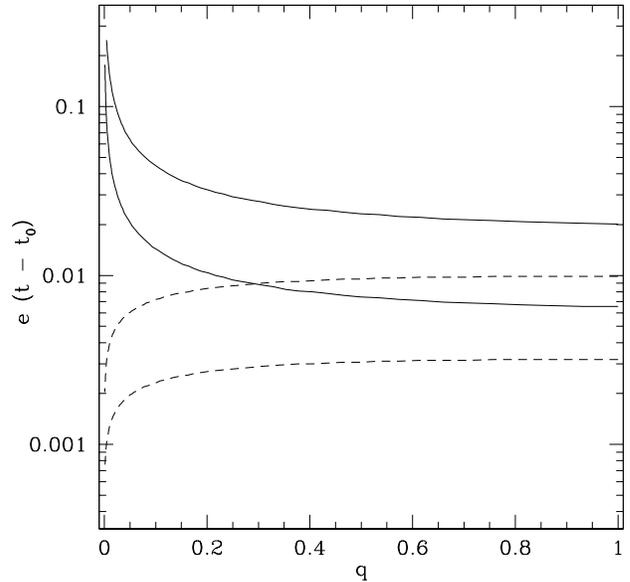,width=\columnwidth}
\figcaption{The predicted binary eccentricity at 1~year (upper solid curve) 
and 1~week (lower solid curve) prior to coalescence as a function of 
mass ratio $q$. We assume that $\dot{m} = 1$, $\alpha = 0.01$, set 
the initial (gas-driven) eccentricity to $e=0.25$, and the total system 
mass $M_1 + M_2 = 10^6 M_\odot$. With these parameters, the transition 
between disk-driven and gravitational wave-driven orbital decay occurs 
close to the transition radius in the disk between radiation and gas pressure 
domination (for $q=1$), and moves to smaller radii for $q \ll 1$. Since 
the migration rate in the radiation pressure dominated disk is uncertain, 
we also plot a lower limit to the final eccentricity (dashed curves), obtained 
by assuming that gravitational radiation always dominates in the radiation 
pressure dominated disk. The dashed curves are evaluated at 1~year and 1~week 
prior to merger.}
\label{fig_eccentric}
\end{figure}

The solid curves in Figure~5 show our best estimates for the estimated eccentricity 
at one week and one year prior to coalescence. We take $\alpha = 0.01$, 
$\dot{m} = 1$, and 
consider binaries of varying $q$ but fixed total mass $M_1 + M_2 = 10^6 M_\odot$. 
Calculations suggest that LISA should detect a significant number of 
mergers in which the black holes have masses in this range \citep{sesana05}, 
whereas events involving more massive holes ($M \sim 10^8 \ M_\odot$) will 
be rare. As previously, we assume that $e=0.25$ at the epoch when gravitational 
radiation becomes the dominant angular momentum loss mechanism and the 
binary decouples from the gas. With these assumptions, we find 
that for roughly equal mass binaries $e \simeq 7 \times 10^{-3}$ at 
one week before coalescence, while a year earlier $e \simeq 0.02$. 
Higher terminal eccentricities are predicted for more extreme 
mass ratios, first because gravitational radiation is relatively 
less efficient, and second because the slowdown in disk driven 
inflow at small radii is smaller for a less massive secondary 
(equation~\ref{eq_slowdown}). The eccentricity at one year 
before coalescence could exceed $e=0.1$ for $q \lesssim 0.02$. 
We also note that there is a substantial separation between the 
decoupling radius -- when the gas ceases to be dynamically 
important for the evolution of the binary -- and the detachment 
radius when the gas and the binary become physically separated.

With these parameters the derived decoupling radius $a_{\rm crit}$ 
moves from modestly interior to $r_{\rm t}$ (for $q=1$), to well 
inside the radiation pressure dominated disk (for $q \ll 1$). The 
uncertainties in calculating $\dot{a}_{\rm inspiral}$ in this 
region therefore have a direct impact on the derived values of 
$a_{\rm crit}$, and on the final estimated eccentricity. Cognizant 
of this, we also plot in Figure~5 a more robust lower limit 
to the final eccentricity, derived under the simple assumption that the 
disk dominates the loss of angular momentum up to the point 
where it becomes radiation pressure dominated.  
We assume an unperturbed disk structure when determining this 
radius (equation~\ref{eq_transition}), so there is no dependence 
on $q$. For $q > 0.2$, the resulting lower limit is approximately 
$e \simeq 0.01$ at one year before coalescence, and $e \simeq 3 \times 10^{-3}$ 
at one week before coalescence. Since the decoupling radius for 
this calculation is fixed, the trend is for smaller final eccentricity 
for lower $q$ (e.g. Figure~4).
 
\section{Discussion and summary} 

In this paper, we have argued that observation of the gravitational 
waves from merging black hole binaries could provide circumstantial 
evidence of the uncertain astrophysical processes that 
(probably) allow supermassive black hole binaries to coalesce. 
Specifically, we suggest that a small but non-zero 
eccentricity would signal that angular momentum loss to gas was 
the dominant decay process immediately preceding the final gravitational
wave-driven inspiral. If, alternatively, interaction with stars 
precedes the final inspiral, the final eccentricity is likely to be 
very small. Most studies suggest that orbital decay via 
stellar dynamical processes is accompanied by circularization 
\citep{polnarev94,makino97,quinlan97}, although \citet{aarseth03} 
has reported eccentricity growth in some recent N-body simulations.

In the case of gas-driven mergers, eccentricity growth appears 
to be a robust prediction of simulations \citep{artymowicz91,
papaloizou01} provided that (a) the mass ratio is not {\em very} 
extreme ($q \sim 0.1$ appears to be safely in the `growth' regime), 
and (b) the surrounding gas is not so hot and dense as to completely 
swamp the binary and thereby fill in the gap whose existence 
is necessary for eccentricity growth \citep{escala04}. One caveat 
is that existing simulations do not treat the angular momentum 
transport within the disk from first principles, and it is 
possible that this omission might alter the conditions needed
for eccentricity growth. Direct simulations of binaries 
embedded within MHD turbulent disks are computationally 
possible (although still difficult) only in the gas pressure 
dominated regime \citep{nelson03,winters03}, and would 
be valuable. There is even more uncertainty as to 
how the binary interacts with a radiation pressure dominated 
disk. 

The presence of significant eccentricity for black hole binaries 
close to merger has additional consequences that are potentially 
observable. As noted by \citet{artymowicz96} both in general, 
and for the specific case of the 
quasar OJ~287, accretion from the circumbinary disk on to the 
individual disks around the holes is strongly time-variable if 
the binary is eccentric. The time scales can be interesting even 
when the binary is still far from coalescence. For example, a 
circumbinary disk surrounding a binary with $M_1 + M_2 = 10^6 M_\odot$ 
and $q=1/4$ would remain gas pressure dominated until the 
separation was less than $\approx 860 r_s$, at which point 
we would predict the disk torque to dominate and the 
eccentricity to be significant. The inspiral time for these 
parameters is of the order of $10^5$~yr, but the orbital 
time scale is still small -- approximately 1~month. This 
is also the time scale on which the individual disks might be 
expected to vary. If gas is responsible for merging a 
large fraction of supermassive black hole binaries, such 
systems, because they are that much further away from merger, 
must be much more numerous than galaxies that will host 
LISA events. Although still rare, they might be 
identified via their characteristic X-ray variability.

The eccentricity is one of the intrinsic parameters of
merging black holes that can be measured via detection 
of the gravitational wave emission, potentially to high 
accuracy. As an illustration, for a {\em stellar mass} 
black hole merging with a $10^6 \ M_\odot$ supermassive 
black hole from an orbit with $e=0.1$, LISA may achieve an 
accuracy of $\Delta e \sim 10^{-4}$ for an event with 
signal-to-noise ratio of 30 \citep{barack04}. \citet{miller05} 
discuss how LISA measurements of the eccentricity of 
compact objects merging with supermassive black holes can 
be used to distinguish between different capture 
mechanisms. Similar measurements 
for mergers of two supermassive holes may 
allow the role of gas in the evolution of binary black 
holes to be studied as a function of the masses of the 
holes involved.   

\acknowledgements

We thank Martin Rees and an anonymous referee for important suggestions -- in 
particular for alerting us to the need to consider the limit of a massive 
secondary -- and acknowledge useful discussions with Mitch Begelman, Paolo 
Coppi, Kristen Menou, Milos Milosavljevic and Mateusz Ruszkowski. 
Parts of this work were supported by NASA under grants NAG5-13207, NNG04GL01G 
and NNG05GI92G from the Origins of Solar Systems, Astrophysics Theory, and 
Beyond Einstein Foundation Science Programs, and 
by the NSF under grant AST~0407040. PJA acknowledges the hospitality of 
the Kavli Institute for Theoretical Physics, supported in part
by the NSF under grant PHY99-07949.

\end{document}